\pdfoutput=1
\documentclass[]{spie}
\usepackage[]{graphicx}
\usepackage{subfigure}
\usepackage{epsfig}

\usepackage[margin=0.5in,bottom=2in,top=1in]{geometry}
\usepackage{lastpage}
\usepackage{fancyhdr}
\usepackage[]{algorithm2e}
\usepackage{amsmath}
\usepackage{tabularx}

\usepackage{tikz}
\usetikzlibrary{shapes,arrows} 
\usepackage{hyperref}
\hypersetup{
    colorlinks,%
    citecolor=blue,%
    filecolor=blue,%
    linkcolor=blue,%
    urlcolor=blue
}

\let\useblackboard=\iftrue
\useblackboard
\font\blackboard=msbm10 scaled \magstep1 \font\blackboards=msbm7
\font\blackboardss=msbm5
\newfam\black
\textfont\black=\blackboard \scriptfont\black=\blackboards
\scriptscriptfont\black=\blackboardss

\else

\fi

\newcommand{\ben}{\begin{eqnarray}\displaystyle}
\newcommand{\een}{\end{eqnarray}}

\tikzstyle{decision} = [diamond, draw, fill=blue!20,  
    text width=8em, text badly centered, node distance=3cm, inner sep=0pt] 
\tikzstyle{block} = [rectangle, draw, fill=blue!20,  
    text width=8em, text centered, rounded corners, minimum height=4em, node distance=4cm] 
\tikzstyle{wideBlock} = [rectangle, draw, fill=blue!20,  
    text width=12em, text centered, rounded corners, minimum height=4em, node distance=4cm] 
\tikzstyle{autoBlock} = [rectangle, draw, fill=blue!20,  
   text centered, rounded corners, minimum height=4em, node distance=4cm] 
\tikzstyle{line} = [draw, -latex'] 
\tikzstyle{cloud} = [draw, ellipse,fill=red!20, 
    minimum height=4em, text width=8em, text centered, node distance=4.5cm] 
\tikzstyle{miniCloud} = [draw, ellipse,fill=red!20, 
    minimum height=4em, text width=5em, text centered, node distance=4.5cm] 
\tikzstyle{autoCloud} = [draw, ellipse,fill=red!20, 
    minimum height=4em, text centered, node distance=4.5cm] 
\tikzstyle{wideCloud} = [draw, ellipse,fill=red!20, 
    minimum height=4em, text width=14em, text centered, node distance=4.5cm] 
\tikzstyle{textBlock} = [ node distance=2cm, 
    minimum height=2em]     
\tikzstyle{imgBlock} = [rectangle, draw, fill=blue!20,  
     minimum height=4em, minimum width=4em] 

\begin{document}

{\center \LARGE \textbf{\textit{Scout-It}:   Interior tomography using modified scout acquisition}}
\linebreak
\linebreak
\vspace{1.5 cm} 
{\Large \textit{Kriti Sen Sharma} }

\tableofcontents
\setcounter{tocdepth}{0}


\section{Introduction}
\label{sec:introduction}

Truncated projections in computed tomography (CT) arise when the patient / object being scanned does not fit within the x-ray beam. This leads to severe artifacts (e.g. cupping, dc bias) in the reconstructed image 	\cite{Natterer2001}. There are several methods that attempt to reduce such `interior reconstruction' artifacts. The first among those methods are those that do not use any prior information e.g. sinogram extension techniques \cite{Kolditz2011,VanGompel2009}, and compressed sensing based reconstruction \cite{Yu2009}. Among these, the former has minimal computational burden but the reconstruction may not be accurate; the latter is highly time consuming and thus may not be commercially viable in many applications. Next, there are  methods that utilize some form of prior information e.g. Hilbert tranform based methods \cite{Noo2004,Defrise2006} that deal with only a certain class of truncated projections, and scout view based methods \cite{SenSharma2013,Xia2015}. 

Scout views are typically always acquired as part of the setup before the final tomographic scan, and thus can be a reliable source of prior information. This author had previously demonstrated that few global scout views can provide highly accurate interior reconstruction in the field of high-resolution micro-CT \cite{SenSharma2013}. More recently, Xia et al. \cite{Xia2015} reported a method to improve image quality in truncated volume-of-interest (VOI) imaging using anterior-posterior (AP) and medio-lateral (ML) scout views. Their work was applied to C-arm based imaging, and relied on the assumption that the scout views cover the entire object, while the final VOI scan is severely truncated. 

In clinical computed tomography (CT), it is quite common that the final tomographic scan is partially truncated (i.e. a fraction of the views are truncated while the rest of the views are non-truncated. In such cases, the AP (anterio-posterior) scout view is usually truncated, but the ML (medio-lateral, or simply `lateral') scout view is not truncated. This paper shows that acquiring AP scout views with a modified configuration allows non-truncated AP scout views. The two non-truncated scout views are then used for a rough estimation of the patient habitus, that in turn allows accurate interior tomography for a truncated tomographic CT scan. 

\section{Methods}
\label{sec:methods}


\subsection{Modified scout configuration}
\label{ssec:modified_scout}

\begin{figure}[hbtp]
\centering
\includegraphics[width=15 cm]{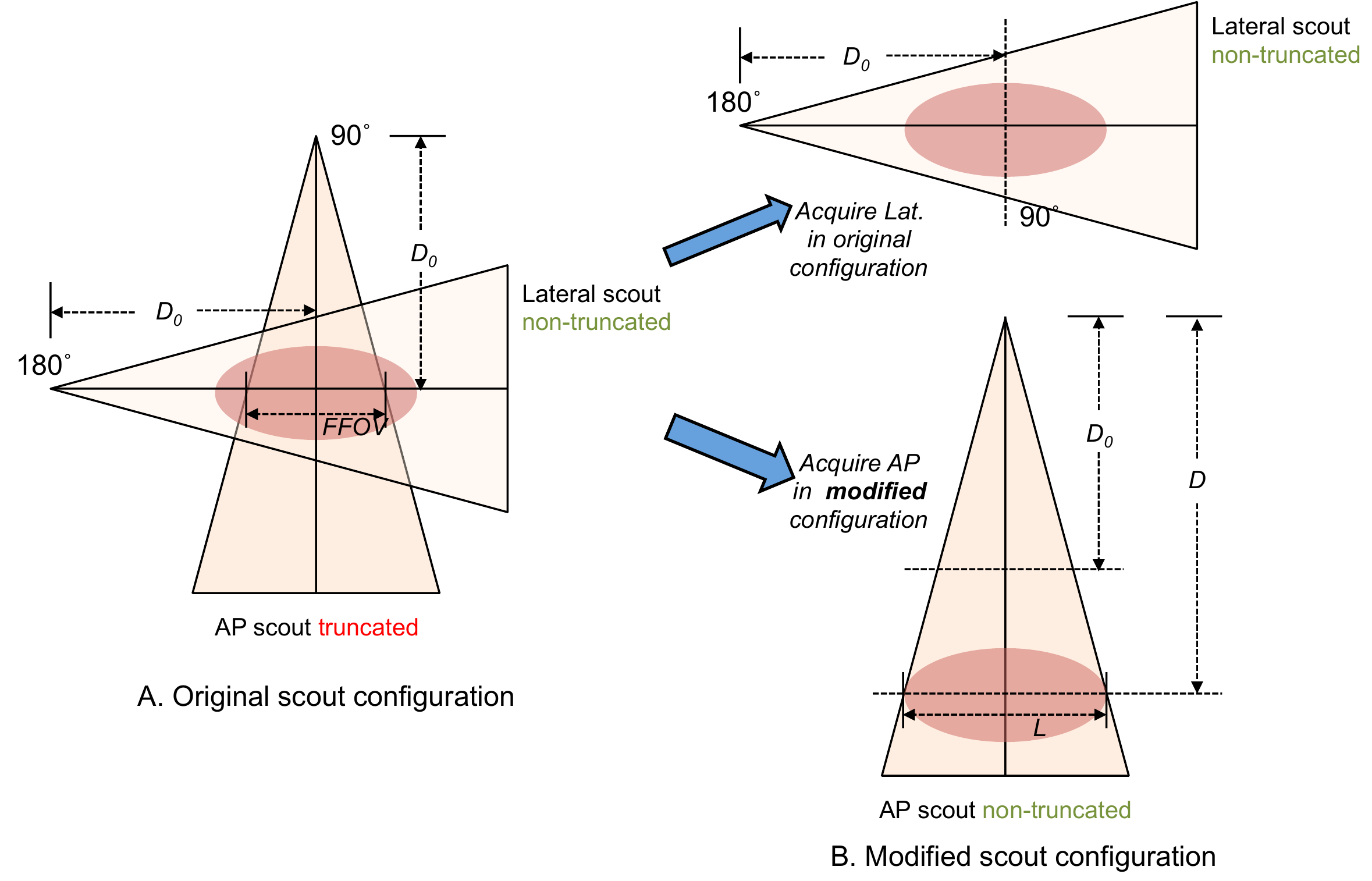}
\caption{Various scout configurations: (A) normal configuration where lateral scout is non-truncated but AP scout is truncated, (B) modified scout configuration where lateral scout is unchanged, but change in source-to-patient distance allows non-truncated AP scout. \label{fig:scout_configurations}
}
\end{figure}

Figure \ref{fig:scout_configurations}A shows a normal scout configuration. The technologist typically attempts to align the patient at the iso-center (patient centering might not be accurate, and is typically improved through information gathered through the scouts). The source-to-(assumed)-iso-center distance is noted as $D_0$ in the figure. For  patients whose body-habitus does not lie within the FFOV of the scanner, the truncation predominantly arises in the AP direction but the lateral (ML) scouts are not truncated. 

As shown in Fig. \ref{fig:scout_configurations}B, we propose a modified scout acquisition. While the lateral (ML) scouts are acquired in the original configuration, the AP scouts are acquired after moving the table down by a distance $d$ (there is of course a limit to how far the table can move down in the gantry). By moving the patient table down, the geometric magnification at the detector changes, and a larger patient habitus can be covered by the x-ray beam. The modified AP scout configuration allows acquisition of non-truncated AP projections up to an extended body habitus of length $L_{\mathrm{eff}}$, where:
\begin{align}
L_{\mathrm{eff}} = \frac{D}{D_0}\times FFOV
\end{align}

We calculate the value for $L_{\mathrm{eff}}$ for four CT scanners: GE LightSpeed Xtra (General Electric Co., Waukesha, USA), Philips Brilliance Big Bore (Koninklijke Philips N.V., Amsterdam, Netherlands), Siemens SOMATOM Sensation Open (24/40) (Siemens AG, Berlin, Germany), and Toshiba Aquilion LB (Toshiba Medical Systems, Otawara, Japan). Geometric specifications for these scanners were retrieved from a report compiled by the ImPACT group, London, UK \cite{ImPACTCenterforEvidencebasedPurchasing2009}. 

In our calculations, we make the following assumptions:
\begin{itemize}
	\item the AP scouts are typically acquired at or near the actual iso-center
	\item the patient table can be moved down in the gantry by at least 150 mm; this appears to be a realistic estimate as the bore radius is greater than 400 mm for all the scanners in list below
\end{itemize}

Table \ref{tab:scanner_geoms} provides a list of relevant geometric specifications reproduced in the ImPACT group report  \cite{ImPACTCenterforEvidencebasedPurchasing2009}, the resulting value of $L_{\mathrm{eff}}$ for a table displacement $d = 150 \, mm$, and the resulting percentage increase (w.r.t. maximum FOV reported by the manufactureer) in patient habitus coverage. The percentage increase lies between 21-26 \%, thus it is evident that the modified scout configuration allows a significant increase in field of view in the AP scout view. 

Next, we show how to combine the original lateral scout and the modified AP scout to reduce truncation artifacts in the final computed tomography scan. The first step (Sec. \ref{ssec:ellipse-estimation}) of the process requires estimation of an ellipse approximately describing the patient habitus. The ellipse estimation uses the information from the original lateral (ML) and modified AP scouts. Next, the estimation can be employed in multiple ways to reduce interior reconstruction artifacts in the reconstruction of the final tomographic scan. Here, we show a simple example using simulations on the Shepp-Logan phantom and the Edge-Gauss projection completion technique \cite{Hoppe2008} (Sec. \ref{sec:results}). 

\begin{table}
\begin{center}
\begin{tabular}{l cccc}
\hline 
\hline 
 & \textbf{GE LSX} & \textbf{Philips Br.} &\textbf{ Siemens SOM.}  & \textbf{Toshiba Aq.}  \\ 
\hline 
Aperture [cm]  & 80 & 85 & 82 & 90 \\ 
Focus-isocentre distance [mm]  & 606 & 645 & 570 & 712 \\ 
Focus-detector distance [mm]  & 1062.5 & 1183 & 1040 & 1275 \\ 
Maximum reconstruction field of view [cm] & 50 & 50 & 50 & 50 \\ 
Reconstruction matrices  & $512 \times 512$ & $512 \times 512$ & $512 \times 512$ & $512 \times 512$ \\ 
\hline 
$L_{\mathrm{eff}}$ [cm] & 62.37 & 61.63 & 63.15 & 60.53 \\ 
\% increase & 24.75 & 23.26 & 26.32 & 21.07 \\ 
\hline 
\hline 
\end{tabular} 
\caption{Geometric specifications of four CT scanners and corresponding values of increased patient coverage $L_{\mathrm{eff}}$. The four CT scanners are GE LightSpeed Xtra (LSX), Philips Brilliance Big Bore (Br.), Siemens SOMATOM Sensation Open (SOM.), Toshiba Aquilion LB (Aq.) \label{tab:scanner_geoms}}
\end{center}
\end{table}
\subsection{Patient shape model estimation using slice-wise ellipse}
\label{ssec:ellipse-estimation}


%
\begin{figure}[hbtp]
\centering
\includegraphics[width=15 cm]{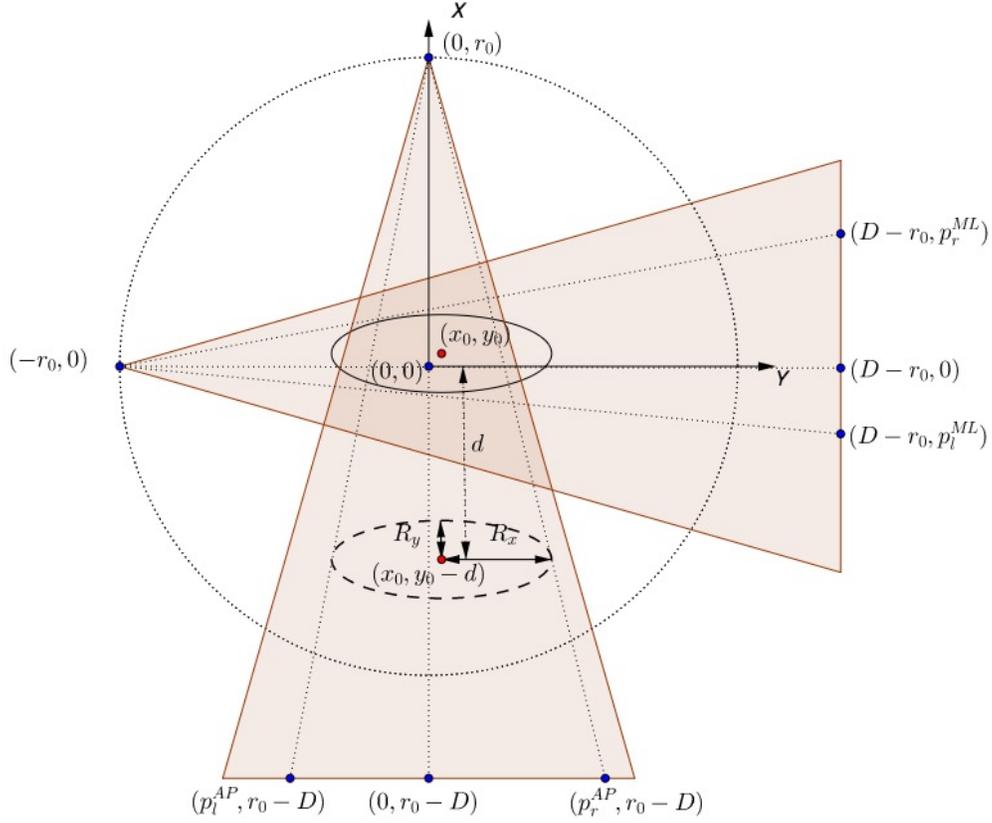}
\caption{Setup for acquisition of ML scout  in original configuration and AP scout in modified configuration \label{fig:scoutIT_geometry}
}
\end{figure}

In our setup shown in Fig. \ref{fig:scoutIT_geometry}, we place the origin of our co-ordinate axes $\left( 0, 0 \right)$ at the scanner iso-center. The source-to-iso-center distance is $r_0$ and the source-to-detector distance is $D$. Thus, the source is at $\left( 0, r_0 \right)$ when acquiring the AP view, and at $\left( -r_0, 0 \right)$ when acquiring the ML view. The patient's body habitus in the slice represented in this diagram is estimated as an ellipse with center $\left( x_0, y_0 \right)$ and major and minor axis lengths of $\left( R_X, R_y \right)$ respectively. 

As shown in the figure, the lateral (ML) scout is non-truncated in the original acquisition configuration. However, the AP scout would be truncated in this original configuration. By lowering the patient table by a distance $d$, the truncation is avoided. The projection of the iso-center, and tangents to the ellipse in the original ML scout and the modified AP scout are marked in the figure. In practice, these values are available from measurements on the acquired scout views. 

We calculate the slopes of the four tangents to the ellipse in ML and modified AP scout views as:
\begin{align}
m_l^{AP} &= \frac{-D}{p_l^{AP}} \nonumber \\
m_r^{AP} &= \frac{-D}{p_r^{AP}} \nonumber \\
m_l^{ML} &= \frac{p_l^{ML}}{D} \nonumber \\
m_r^{Ml} &= \frac{p_r^{ML}}{D} 
\end{align}
where the suffix $p_l$ and $p_r$ denote the left and right tangents. 

Next, we note the equation of the tangents below.

Tangents in ML scout:
\begin{equation}
\frac{y-0}{x-\left(-r_0 \right) } = m^{ML} \label{eqn:tangent_mlscout}
\end{equation}

Tangents in AP scout:
\begin{equation}
\frac{y-r_0}{x-0} = m^{AP} \label{eqn:tangent_apscout}
\end{equation}

The equation of the ellipse in the two configurations are noted below.

Ellipse in ML scout:
\begin{equation}
\left( \frac{y - y_0 }{R_y} \right)^2 + \left( \frac{x-x_0}{R_x}\right)^2 = 1 \label{eqn:ellipse_mlscout}
\end{equation}

Ellipse in modified AP scout configuration:
\begin{equation}
\left( \frac{y - \left( y_0 - d \right) }{R_y} \right)^2 + \left( \frac{x-x_0}{R_x}\right)^2 = 1 \label{eqn:ellipse_apscout}
\end{equation}

Now, the system of equations provided by equations  \ref{eqn:ellipse_mlscout} and \ref{eqn:tangent_mlscout} allow us to solve for the points of intersection of tangents to the ellipse in the ML scout. We apply the following operations on the system of equations:
\begin{enumerate}
	\item Though the value of $m_ML$ is known, we do not substitute this into the equation at first
	\item Substitute the equation for $y$ provided in Eq. \ref{eqn:tangent_mlscout} into Eq. \ref{eqn:ellipse_mlscout} \label{algo:subst1}
	\item Step \ref{algo:subst1} yields a quadratic in $x$. Since the the system of equation allows only a unique solution for each tangent, we can apply the formula that the discriminant of this quadratic is zero	 
\label{algo:discr}
	\item Step \ref{algo:discr} yields a quadratic in $m$. The parameters of the ellipse $\left\lbrace x_0, y_0, R_x, R_y \right\rbrace$ are also embedded in this equation
	\item Using the above quadratic which is of the form $A m^2 + B^m + C = 0$, we use the formulae for sum of roots and difference of roots
\end{enumerate}

By following the above set of steps for both the ML and AP scouts, we arrive at the following system of equations:
\begin{align}
\Sigma_1 \left( B x_0^2 - 1 \right) + 2 B x_0 \left( r_0 + d - y_0 \right) = 0  \nonumber \\
A \Delta_1^2 \left(B x^2 - 1 \right) - 4 B \left(A \left( r_0 + d - y_0 \right) ^2 + B x_0^2 - 1 \right) = 0 \nonumber \\
\Sigma_2 \left( B \left( x_0 + r_0 \right) ^2 - 1 \right) + 2 B y_0 \left( x_0 + r_0 \right) = 0  \nonumber  \\
A \Delta_2^2 \left(B \left( x_0 + r_0 \right) ^2 - 1 \right) - 4 B \left(A y_0^2 + B \left( x_0 + r_0 \right) ^2 - 1 \right) = 0 
\label{eq:finalsysofeqs}
\end{align}
where we used the following substitutions to simplify the equations:
\begin{align}
A = \frac{1}{R_y^2}  \nonumber  \\
B = \frac{1}{R_x^2}  \nonumber  \\
\Sigma_1 = m_l^{AP} + m_r^{AP} \nonumber  \\
\Sigma_2 = m_l^{ML} + m_r^{ML} \nonumber  \\
\Delta_1 = m_l^{AP} - m_r^{AP} \nonumber  \\
\Delta_2 = m_l^{ML} - m_r^{ML}
\end{align}

Thus we have a system of four equations with four unknowns $\left\lbrace x_0, y_0, R_x, R_y \right\rbrace$. This system can be solved with numerical solvers to yield the parameters of the ellipse (i.e. an estimate of the patient body habitus). 

\section{Results}
\label{sec:results}

\subsection{Numerical phantom}
The proposed method is tested through numerical simulation. The Shepp-Logan phantom was expressed into a $512 \times 512$ image grid. At this image grid, the major and minor axes of the major ellipse comprising the Shepp Logan phantom were measured manually as $R_x = 233$ and $R_y = 177$ pixels. When simulations are carried out on the Shepp-Logan phantom, the major ellipse is typically centered at the iso-center of the simulated gantry. For the purpose of this paper (and in most real-life situations), this is not a realistic assumption because the patient is not always centered perfectly at the gantry iso-center. Keeping this in mind, we assumed for our tests that the phantom was not perfectly centered at the gantry iso-center. In the following test, we used a perturbation from iso-center $(x_0 = -5, y_0 = -8)$ pixels (i.e. a 2.1\% deviation w.r.t. $R_x$ and a 4.5\% deviation w.r.t $R_y$). While these values were arbitrarily selected, the accuracy of the proposed method is expected to hold for other values of perturbation from gantry iso-center (provided $x_0$ and $y_0$ are within reasonable bounds of patient centering error). 

\subsection{Projection acquisition}
\begin{figure}[hbtp]
\centering
\includegraphics[width=15 cm]{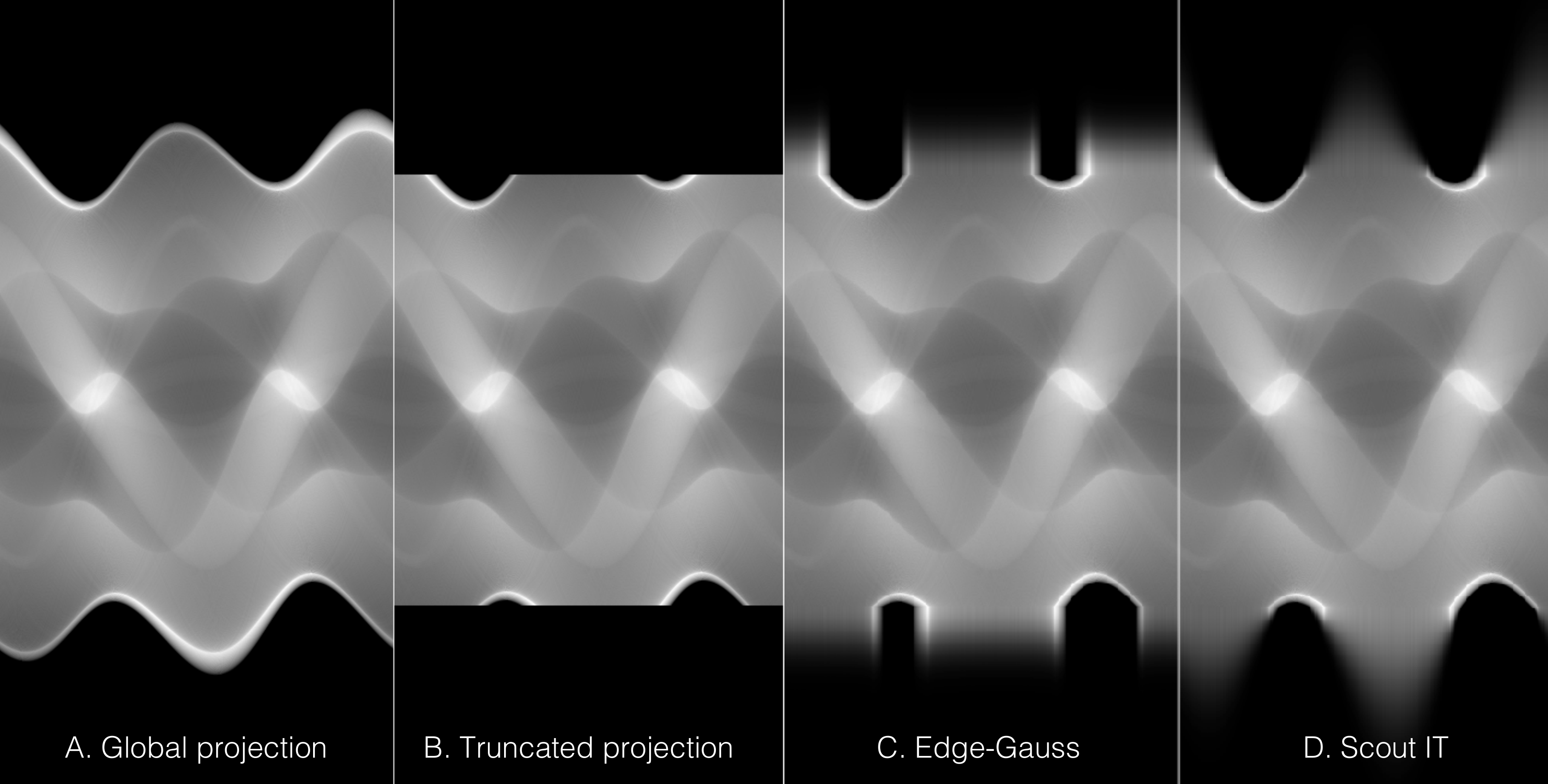}
\caption{Sinograms used in the study: (A) Global projection sinogram (reference), (B) Truncated projection (note that some of the ML views are non-truncated, (C) Projection completion by Edge-Gauss method without knowledge of patient boundary information, (D) Projection completion by Edge-Gauss method utilizing patient boundary estimated by proposed method. \label{fig:sinograms}
}
\end{figure}

Next, scout projections and entire tomographic scan were generated for the phantom. The scan parameters used in our test (along with the code) are provided at \url{https://github.com/ksens/scoutIT}. As seen in Fig. \ref{fig:sinograms}B, the AP projections are truncated, but the ML projections are not. Similarly, the AP scout is truncated, but the ML scout is non-truncated in this configuration. By moving the detector down by 200 mm, a non-truncated AP scout was acquired. As a reference, the global tomographic scan is shown in Fig. \ref{fig:sinograms}A.  


\subsection{Estimation of ellipse parameters}

The edges of the non-truncated scout views were measured and the slope of the tangents were calculated using equations \ref{eqn:tangent_mlscout} and \ref{eqn:tangent_apscout}. Finally, the system of equations \ref{eq:finalsysofeqs} was solved to determine the unknowns $x_0, y_0, R_x$ and $R_y$. The results are tabulated in Table \ref{tab:ellipse_params}; the derived values match closely with the values of ellipse parameters used in the simulation. 

\begin{figure}[hbtp]
\centering
\includegraphics[width=15 cm]{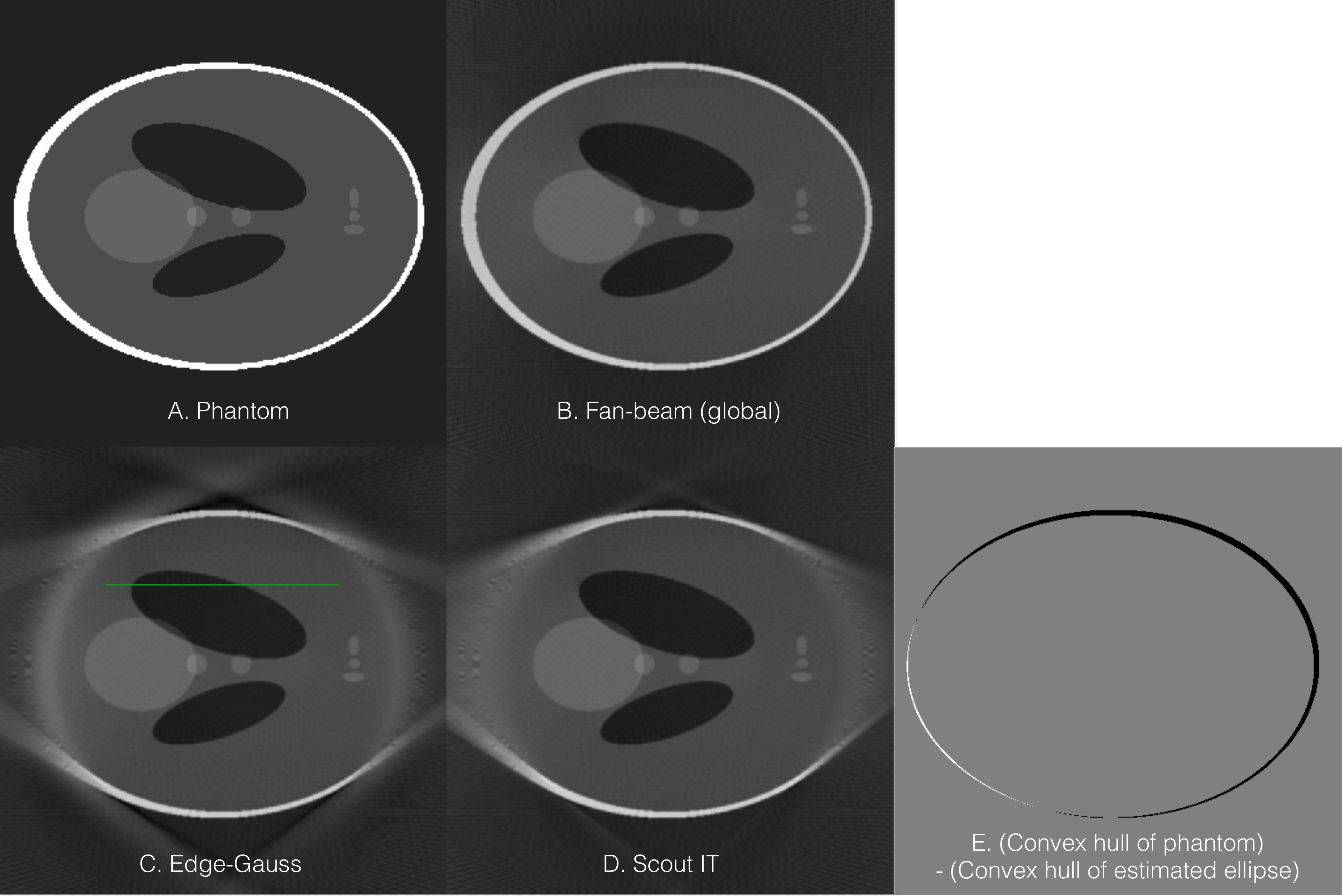}
\caption{Images of Shepp-Logan phantom: (A) Original phantom, (B) Global fan-beam reconstruction (shown as a reference), (C) Reconstruction of projection completed by Edge-Gauss method without patient boundary information, (D) Reconstruction of projection completed by proposed method, (E) Difference image of convex hull of phantom, and estimated ellipse. \label{fig:recon-images}}
\end{figure}

As further evidence, the difference image between the binary maps of the convex hull of the Shepp Logan phantom and the estimated ellipse is shown in Fig. \ref{fig:recon-images}E. 

\begin{table}
\begin{center}
\begin{tabular}{|c|c|c|c|c|}
\hline 
 & \textbf{$R_x$} & \textbf{$R_y$} & \textbf{$x_0$} & \textbf{$y_0$} \\ 
\hline 
Actual (pixels) & 233 & 177 & -5 & -8 \\ 
\hline 
Proposed method (pxls) & 233 & 173 & -4.89 & -8.16 \\ 
\hline 
\end{tabular} 
\caption{\label{tab:ellipse_params} Actual ellipse parameters used in the numerical simulation (first row), and results of the proposed method (second row).}
\end{center}
\end{table}

\subsection{Usage of ellipse parameters in interior tomography}

\begin{figure}[hbtp]
\centering
\includegraphics[width=15 cm]{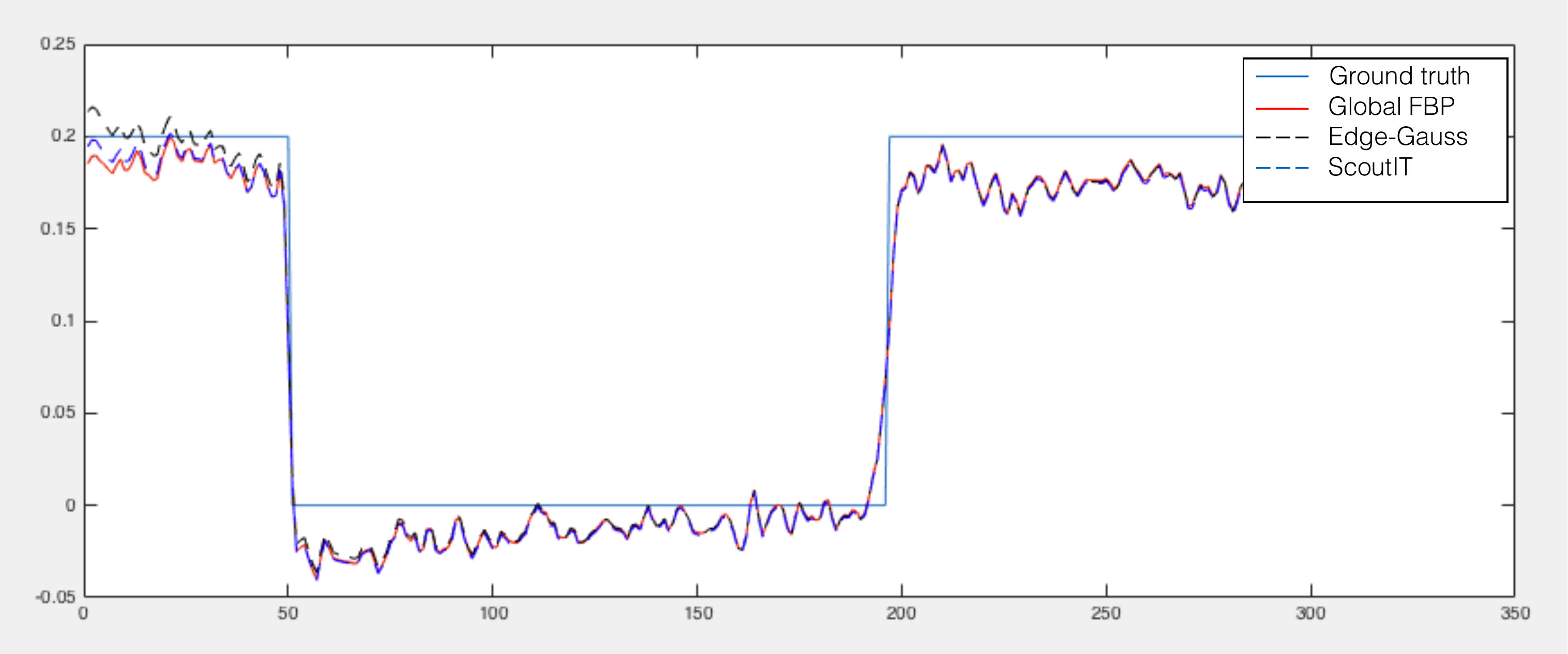}
\caption{Line profiles across segment marked in Fig. \ref{fig:recon-images}: (A) Original phantom, (B) Global fan-beam reconstruction (shown as a reference), (C) Reconstruction of projection completed by Edge-Gauss method without patient boundary information, (D) Reconstruction of projection completed by proposed method \label{fig:lineprofiles}}
\end{figure}

For reconstruction of the truncated sinogram, we utilized the simple (yet popular) method of Edge-Gauss extrapolation of truncated projections \cite{Hoppe2008}. As a reference, we first ran extrapolation without knowledge of the object boundary. This kind of extrapolation is commonly employed in the field, and the resultant extrapolated sinogram is shown in Fig. \ref{fig:sinograms}C. The image reconstructed by this method is shown in Fig. \ref{fig:recon-images}C. Line profile is taken across the green segment marked on this figure, and plotted in Fig. \ref{fig:lineprofiles}. The reconstruction is quite similar to the global fan-beam reconstruction -- however, the left portion of the line profile shows that the reconstruction value differs significantly from the global fan-beam reconstruction (the global fan-beam reconstruction being the best-case achievable result). 

Finally, we utilize the derived values for the ellipse parameters ($x_0, y_0, R_x, R_y$) for accurate interior tomography. The estimated ellipse provides the additional knowledge about the patient habitus. This information is utilized while employing projection completion using Edge-Gauss technique. The projection completed sinogram by proposed method is shown in Fig. \ref{fig:sinograms}D; clearly, there is a much greater resemblance with the global projection in this case. The reconstructed image is shown in Fig. \ref{fig:recon-images}D, and the line profile is shown in Fig. \ref{fig:lineprofiles}. The proposed method allows the final reconstruction to be very similar to global fan-beam reconstruction, even in the left-portion of the line-profile. 

\section{Discussion}

We showed that, for Shepp-Logan phantom placed at an arbitrary displacement from iso-center, if ML scout is non-truncated and AP scout in modified configuration is non-truncated, then the four parameters of the ellipse can be accurately determined. Next, we showed that knowledge of the patient habitus (approximated by an ellipse in our case) leads to more accurate interior tomography than existing methods (that do not use knowledge acquired from the scout views). Futures steps include demonstration of the proposed method for truncated clinical projection data.
\bibliography{bibl_scoutit}   
\bibliographystyle{spiebib}   

\end{document}